# Fast, low-current spin-orbit torque switching of magnetic tunnel junctions through atomic modifications of the free layer interfaces

Shengjie Shi[1], Yongxi Ou[1], S.V. Aradhya[1], D.C. Ralph[1,2] and R.A. Buhrman[1*]

[1]Cornell University, Ithaca, New York 14853, USA

[2]Kavli Institute, Cornell University, Ithaca, New York 14853, USA

**Future applications of spin-orbit torque will require new mechanisms to improve the efficiency for switching nanoscale magnetic tunnel junctions (MTJs), while also controlling the magnetic dynamics to achieve fast, nanosecond scale performance with low write error rates. Here we demonstrate a strategy to simultaneously enhance the interfacial magnetic anisotropy energy and suppress interfacial spin memory loss by introducing sub-atomic and monatomic layers of Hf at the top and bottom interfaces of the ferromagnetic free layer of an in-plane magnetized three-terminal MTJ device. When combined with a beta-W spin Hall channel that generates spin-orbit torque, the cumulative effect is a switching current density of 5.4 x $10^6$ A/cm$^2$, more than a factor of 3 lower than demonstrated in any other spin-orbit-torque magnetic memory device at room temperature, and highly reliable switching with current pulses only 2 ns long.**

*buhrman@cornell.edu.

Spin-orbit torque (SOT) from the spin Hall effect (SHE) [1–8] in heavy metals (HMs) can rapidly and reliably switch an adjacent ferromagnet (FM) free layer of a nanoscale MTJ in a three-terminal configuration (3T-MTJ). This effect provides the strategy for a new generation of fast, current- and energy-efficient cache magnetic memory [9–15]. The separate read and write channels in the 3T-MTJ geometry offer additional advantages; faster read-out without read disturbance and lower write energy. While the development of SOT switching has focused primarily on nanoscale perpendicularly magnetized MTJs, their SOT effective-field switching requires much higher currents than can be provided by a reasonably scaled CMOS transistor (current densities in the SH channel are $\geq 1.4 \times 10^8$ A/cm$^2$) [16], and fast, low write error rate (WER) switching has not yet been demonstrated. Here we report a dramatic performance improvement for *in-plane-magnetized* 3T-MTJs wherein the strong SOT arising from nano-channels of beta-phase W is combined with two recently discovered effects of Hf atomic layer modifications of the FM-MgO and HM-FM interfaces that, respectively, enhance the interfacial perpendicular magnetic anisotropy energy density [17] and reduce interfacial spin memory loss [18]. The result is an anti-damping SOT switching current density of just $5.4 \times 10^6$ A/cm$^2$. We also achieve reliable (write error rate (WER) $\approx 10^{-6}$) switching with 2 ns pulses, which we tentatively attribute, at least in part, to the beneficial assistance of the field-like SOT arising from the spin current generated by the W spin Hall effect.

The high performance 3T-MTJ devices reported here were lithographically patterned from a thin film multilayer stack sputter-deposited onto an oxidized Si wafer consisting of W(4.4)/Hf(0.25)/Fe$_{60}$Co$_{20}$B$_{20}$(1.8)/Hf(0.1)/MgO(1.6)/Fe$_{60}$Co$_{20}$B$_{20}$(4)/Ta(5)/Ru(5) (thickness in nanometers), where W represents the high-resistivity beta-phase of W [19]. In Fig. 1a we show a schematic of the W-based 3T-MTJ device structure along with (inset) a scanning electron microscope (SEM) image of a typical elliptical nano-pillar MTJ on top of the W SHE channel after it has been defined by electron-beam lithography and argon ion milling.

We demonstrate the potential of these W-based in-plane-magnetized (IPM) 3T-MTJ devices by reporting in detail on the representative performance of a high-aspect-ratio, 30nm x 190nm, MTJ device fabricated on a 480nm wide W channel. This device was annealed in an air furnace at 240 C for 1 hour after patterning to increase the tunneling magnetoresistance (TMR) of the MTJ and also reduce the switching current as discussed below. In the inset to Fig. 1b we first show the minor magnetic loop response of the MTJ resistance as an in-plane magnetic field $H_{ext}$ is applied along the long axis of the MTJ device and ramped over ± 300 Oe, which is sufficient to reverse the orientation of the thin bottom free layer (FL) of the MTJ from being parallel (P) to anti-parallel (AP) to the thicker FeCoB reference layer, but not strong enough to reverse the orientation of the reference layer due to its stronger shape anisotropy. The horizontal offset of the minor loop (~ -50 Oe) is due to the dipole field from the reference layer. All subsequent SOT measurements are taken when this offset is canceled by an appropriate $H_{ext}$.

In the main part of Fig. 1b we show the characteristic DC SOT hysteretic switching behavior of the IPM 3T-MTJ as the bias current in the W channel is ramped quasi-statically. The switching polarity is consistent with the negative spin Hall sign of β-W in comparison to that of platinum [4,20]. For nanoscale MTJs thermal fluctuations assist the reversal during slow current ramps. Within the macrospin or rigid monodomain model the critical current $I_c$ that is observed is dependent on the current ramp rate [21],

$$I_c = I_{c0}\left\{1-\frac{1}{\Delta}\ln\left[\frac{1}{t_0\Delta}\left(\frac{|I_{c0}|}{|\dot{I}|}\right)\right]\right\} \qquad (1)$$

Here $I_{c0}$ is the critical current in the absence of thermal fluctuation, $\dot{I}$ is current ramp rate, $\Delta$ is the thermal stability factor that represents the normalized magnetic energy barrier for reversal between the P and AP states, and $\tau_0$ is the thermal attempt time which we assumed to be 1 ns. To characterize the SOT behavior of this device we measured the mean switching current for $\dot{I}$ varying from $10^{-7}$A/s to $10^{-5}$A/s (Fig. 1c). By fitting to Eqn. (1) we obtained nearly symmetric SOT switching results with an averaged zero-fluctuation switching current of $|I_{c0}| = 115$ µA and $\Delta = 35.6$. With the W channel width $w_{SH} = 480$ nm and thickness $t_{SH} = 4.4$ nm this corresponds to a switching current density $J_{c0} = 5.4 \times 10^6 A/cm^2$, more than three times lower than reported originally for a W-based 3T-MTJ [4] and by far the lowest yet reported for any 3T-MTJ device with $\Delta > 35$.

In Table 1 we compare critical switching current density $J_{c0}$ achieved in various in-plane and out-of-plane 3T structures. The different types of SOT devices have different minimum sizes as determined by thermal stability requirements, which in turn will set the current amplitude required for switching or domain wall motion. For the PMA SOT nanodot devices currently a 40 nm diameter is required [16], which would necessitate a minimum current of approximately 300 µA for reversal using a 40 nm wide, 4 nm thick beta-W spin Hall channel. In comparison our in-plane magnetized 3T-MTJ 190nm x 30nm device would require a switching current of approximately 40 µA for a 190 nm wide channel.

The critical question is whether these W-based IPM 3T-MTJ devices can also exhibit fast reliable switching with low amplitude current pulses as required for high-speed cache memory. Aradhya et al. have recently reported nanosecond timescale switching of Pt-based IPM 3T-MTJs with low WER [20], but a high current pulse amplitude was required, 2-3 $I_{c0}$, with $I_{c0}$ > 500 µA. To characterize the performance of the W-based 3T-MTJ device in the short pulse regime, we separately measured the switching phase diagram for the two cases, $P \rightarrow AP$ and $AP \rightarrow P$, using a fast pulse measurement method [19]. Results are shown in Fig. 2a and 2b where each data point is the statistical result of 1000 switching attempts, with the scale bar on the right showing the switching probability. Although micromagnetic modeling indicates that for strong short pulses these 3T-MTJ devices do not reverse simply as a rigid domain [20] we can still utilize the macrospin model [22] as an approximation to characterize the short pulse response by fitting the 50% switching probability boundary between the switching and non-switching regions with,

$$V = V_0(1 + \frac{\tau_0}{t}) \qquad (2)$$

The results shown in the solid curves provide a reasonable fit to the data despite the simplifying macrospin assumption. From these fits we extract the characteristic switching times $\tau_0$ and critical switching voltages $V_0$ to be 0.76ns and 0.48V for $P \rightarrow AP$ and 1.20 ns and 0.44V for $AP \rightarrow P$. The short pulse critical switching current (current density) as calculated from $V_0$ and the channel resistance $R \approx 3.6 k\Omega$ is $I_{c0} \approx 120 \mu A$ ($J_{c0} \approx 5.9 \times 10^6 A/cm^2$), consistent with the ramp rate results.

For cache memory SOT reversal has to be both fast and highly reliable and in this latter regard our results with this W-based IPM 3T-MTJ approach offer encouraging prospects as indicated by Fig. 2c, where we show WER results measured with 2ns pulses on the same device. We applied square switching pulses of increasing voltages to the W channel and recorded states of the device after each switching pulse. For every voltage level we repeated the switching attempts $10^6$ times and calculated WER based on switching probability $WER = 1 - P_{switch}$. At 2ns, WER of close to $10^{-6}$ is achieved for both polarities $P \rightarrow AP$ and $AP \rightarrow P$, which indicates the potential of this approach for high reliability. Note that our current results were limited to $10^{-6}$ WER (V ≤ 3.5 $V_0$) due to the constraint on the highest pulse voltage we can apply to the channel imposed by a less than optimal electrode design (spreading resistance) and a poor quality field insulator. Straightforward improvements in both should lower $V_0$ and enable measurements with V >> $V_0$.

The observed anti-damping SOT reversal on a ≤ 1 ns timescale is much faster than predicted by the rigid domain, macrospin model. A key conclusion in the initial report on fast switching with Pt-based IPM 3T-MTJs was that the in-plane Oersted field $H_{Oe}$ generated by the pulsed current is advantageous in promoting the fast reliable switching because it opposes the anisotropy field $H_c$ of the FL at the beginning of the reversal [20,23]. Due to the opposite sign of the SHE for W-based 3T-MTJs the pulsed $H_{Oe}$ in our case is parallel to $H_c$ at the beginning of the pulse which micromagnetic modeling indicated should be disadvantageous for very fast reversal [20]. However we have made spin-torque ferromagnetic resonance (ST-FMR) measurements on W(4)/Hf(0.25)/FeCoB(/Hf(0.1)/MgO/Ta microstrips that had been annealed at 240C for 1 hour to determine both the *anti-damping* and the *field-like* spin-orbit torque efficiencies, $\xi_{DL}$ and $\xi_{FL}$, of this heterostructure and we obtained $\xi_{DL} = -0.20 \pm 0.03$ and $\xi_{FL} = -0.0364 \pm 0.005$ [19]. This *field-like* torque efficiency corresponds to an effective field $-6.68 \times 10^{-11}$ Oe/(A/m$^2$) in the MTJ structure with a 1.8 nm free layer that is oriented in opposition to the Oersted field generated by the electric current, as previously reported for W devices [4], and approximately three times larger. Thus the net transverse field is in opposition to the free layer in-plane anisotropy field at the beginning of the reversal and hence may be playing an important role in the fast reliable W-based 3T-MTJ results reported here.

In addition to utilizing the high spin torque efficiency of $\beta-W$ we have employed two other materials enhancements, the sub-monolayer "dusting" and monolayer "spacer" of Hf that were inserted respectively between the FL and the MgO and between the W and the FL, to achieve this exceptionally low pulse current (density) switching performance. For 3T-MTJs the SOT switching current density, within the macrospin model, is predicted to vary as [24,25]

$$J_{c0} = I_{c0}/w_{SH}t_{SH} = \frac{2e}{\hbar}\mu_0 M_s t_{FM}\alpha(H_c + M_{eff}/2)/\xi_{DL} \qquad (3)$$

where e is the electron charge, $\hbar$ is the reduced Plank constant, $\mu_0$ is the permeability of free space, $M_s$ is the saturation magnetization of the FL and $t_{FM}$ is the FL's effective magnetic thickness, which were measured to be $1.2\times10^6 A/m$ and 1.7 nm [19], $M_{eff} \equiv M_s - K_s/t_{FM}$ is the FL's effective demagnetization field, where $K_s$ is the interfacial perpendicular magnetic anisotropy energy density, and $\alpha$ is the effective magnetic damping constant of the FL. To compare our experimental results with the prediction of Eqn. (3) we conducted a flip-chip ferromagnetic resonance (FMR) measurement [19] of an un-patterned section of the wafer to determine $M_{eff}$ = 2110 Oe and $\alpha$ = 0.012. With these parameter values we calculate from Eqn. (3) that $\xi_{DL} = 0.15 \pm 0.03$ for the measured device, a bit lower than the result from the ST-FMR measurement of a larger area microstrip of the same heterostructure composition [19]. This difference may be due to an increase in damping due to side-wall oxidation of the nanopillar in the lithography process, which can be addressed by in-situ passivation in the future [26].

The benefits of the Hf insertion layers for reducing the critical current for SOT switching are illustrated by comparisons with FMR measurements performed on two control samples, one with only the Hf dusting,

W(4)/FeCoB(1.8)/Hf(0.1)/MgO(1.6)/FeCoB(4)/Ta(5)/Ru(5), and one without either Hf layer W(4)/FeCoB(1.8)/MgO(1.6)/FeCoB(4)/Ta(5)Ru(5). Consistent with a previous report that Hf dusting can greatly enhance the perpendicular magnetic anisotropy energy density $K_s$ at FM/MgO interfaces [17] we found that $M_{eff}$ for the Hf dusting-only structure was reduced to 4300 Oe, compared to 9860 Oe for the W MTJ system without Hf (Fig. 1d). The additional reduction to $M_{eff}$ = 2110 Oe for the sample with the added Hf spacer can be attributed to some of that Hf diffusing through the FeCoB to the MgO interface during the anneal [18,27]. Another benefit of the Hf spacer is that its insertion decreases $\alpha$ very substantially from 0.018 to 0.012 (Fig. 1e) which we attribute to a passivation of the W surface that suppresses reaction between the W and FeCoB that would otherwise result in interfacial spin memory loss [28]. While there is some spin current attenuation from the use of the Hf spacer [18,29], its effectiveness in lowering the effective damping, and $M_{eff}$ substantially outweighs that cost.

Integration of MRAM with CMOS usually requires thermal processing above 240 C. Annealing at higher temperatures can also be beneficial in producing higher TMR. The 30 nm x 190 nm free layers analyzed above became thermally unstable due to further decrease in $M_{eff}$ after annealing at 300 C, but it is important to note that the Hf dusting technique itself becomes even more effective after processing at T ≥ 300 C. We performed FMR measurements on an un-patterned section of the wafer with only the 0.1nm Hf dusting layer after it was annealed at 300C for one hour. As illustrated in Fig 3a raising the annealing temperature from 240C to 300C resulted in approximately a 2.5x reduction in $M_{eff}$ from 4300 Oe to 1550 Oe, while there was no material effect on $M_s$ [19], a compelling demonstration of the effectiveness of Hf dusting in enhancing $K_s$. To examine the SOT switching behavior of devices with such low $M_{eff}$ we patterned larger 390nm x 100nm, and hence more thermally stable, MTJs from the wafer and annealed two of them at the two different temperatures, 240 C and 300 C respectively. Consistent with the $M_{eff}$ change, we see clean SOT switching with a much lower critical current, $I_{c0} = 155 \mu A$, after 300 C annealing temperature in comparison to the 240 C critical current $I_{c0} = 335 \mu A$.

In summary, we have achieved nanosecond-scale, reliable, low-amplitude pulse current switching in W-based IPM 3T-MTJs by utilizing a partial atomic layer of Hf dusting between the FL and the MgO which very effectively reduces $M_{eff}$ of the FL, while a further reduction in the required pulse amplitude is achieved by inserting approximately one Hf monolayer between HM and FM which significantly reduces interfacial spin memory loss. This ability to achieve a low $M_{eff}$ with a relatively thick free layer through use of the particularly strong interfacial anisotropy effect of Hf-O-Fe bonds [17] has enabled us to minimize the detrimental effect of interfacial enhancement of magnetic damping, and, due to the thicker free layer, arguably also hinders the formation of localized non-uniformities during the fast reversal that would otherwise result in write errors.

Further decreases in $I_c$, to well below 100 µA, should be quite straightforward with refinements in device design. For example, to ensure successful fabrication, the major axis of our elliptical MTJ nanopillars is less than 50% the width of the spin Hall channel so that up to a factor of two reduction in $I_c$ can be expected simply with more aggressive, industry-level lithography. Smaller nanopillars on even narrower channels, ≤ 100 nm, should be possible through the use of slightly thicker FLs to promote thermal stability, with the robust interfacial magnetic anisotropy effect of the Hf dusting technique providing the means to achieve a low $M_{eff}$ even for $t_{FM} \geq 2$ nm. We anticipate that these approaches, in conjunction with an improved device geometry that substantially reduces the spreading resistance, should lower the pulse write current for fast, reliable switching to ≈ 20 µA and the write energy to the ≤ 10 fJ scale.

S.S and Y.O contributed equally to this work.

**Acknowledgements**

The authors thank C.L. Jermain and N.D. Reynolds for assistance with the FMR measurement. This report is based upon work supported by the Office of the Director of National Intelligence (ODNI), Intelligence Advanced Research Projects Activity (IARPA), via contract W911NF-14-C0089. The views and conclusions contained herein are those of the authors and should not be interpreted as necessarily representing the official policies or endorsements, either expressed or implied, of the ODNI, IARPA, or the U.S. Government. The U.S. Government is authorized to reproduce and distribute reprints for Governmental purposes notwithstanding any copyright annotation thereon. Additionally, this work was supported by the NSF/MRSEC program (DMR-1120296) through the Cornell Center for Materials Research, by the Office of Naval Research, and by the NSF (Grant No. ECCS-0335765) through use of the Cornell NanoScale Facility/National Nanotechnology Coordinated Infrastructure.

References

[1]  I. M. Miron, G. Gaudin, S. Auffret, B. Rodmacq, A. Schuhl, S. Pizzini, J. Vogel, and P. Gambardella, Nat. Mater. **9**, 230 (2010).
[2]  L. Liu, C.-F. Pai, Y. Li, H. W. Tseng, D. C. Ralph, and R. A. Buhrman, Science **336**, 555 (2012).
[3]  I. M. Miron, K. Garello, G. Gaudin, P.-J. Zermatten, M. V Costache, S. Auffret, S. Bandiera, B. Rodmacq, A. Schuhl, and P. Gambardella, Nature **476**, 189 (2011).
[4]  C. F. Pai, L. Liu, Y. Li, H. W. Tseng, D. C. Ralph, and R. A. Buhrman, Appl. Phys. Lett. **101**, 122404 (2012).
[5]  S. Fukami, C. Zhang, S. Duttagupta, A. Kurenkov, and H. Ohno, Nat. Mater. **15**, 535 (2016).
[6]  Y. Ou, S. Shi, D. C. Ralph, and R. A. Buhrman, Phys. Rev. B **93**, 220405(R) (2016).
[7]  P. M. Haney, H. W. Lee, K. J. Lee, A. Manchon, and M. D. Stiles, Phys. Rev. B **87**, 174411 (2013).
[8]  G. Finocchio, M. Carpentieri, E. Martinez, and B. Azzerboni, Appl. Phys. Lett. **102**, 212410 (2013).
[9]  N. Locatelli, V. Cros, and J. Grollier, Nat. Mater. **13**, 11 (2014).
[10] A. D. Kent and D. C. Worledge, Nat. Nanotechnol. **10**, 187 (2015).
[11] S. Fukami, T. Anekawa, C. Zhang, and H. Ohno, Nat. Nanotechnol. **11**, 621 (2016).
[12] G. Yu, P. Upadhyaya, Y. Fan, J. G. Alzate, W. Jiang, K. L. Wong, S. Takei, S. a Bender, L.-T. Chang, Y. Jiang, M. Lang, J. Tang, Y. Wang, Y. Tserkovnyak, P. K. Amiri, and K. L. Wang, Nat. Nanotechnol. **9**, 548 (2014).
[13] A. Hoffmann, IEEE Trans. Magn. **49**, 5172 (2013).
[14] S. Ikeda, K. Miura, H. Yamamoto, K. Mizunuma, H. D. Gan, M. Endo, S. Kanai, J. Hayakawa, F. Matsukura, and H. Ohno, Nat. Mater. **9**, 721 (2010).
[15] M. Cubukcu, O. Boulle, M. Drouard, K. Garello, C. O. Avci, I. M. Miron, J.

Langer, B. Ocker, P. Gambardella, G. Gaudin, M. Cubukcu, O. Boulle, M. Drouard, K. Garello, and C. O. Avci, Appl. Phys. Lett. **104**, 42406 (2014).

[16] S. Fukami and H. Ohno, Jpn. J. Appl. Phys **56**, 0802A1 (2017).

[17] Y. Ou, D. C. Ralph, and R. A. Buhrman, Appl. Phys. Lett. **110**, 192403 (2017).

[18] M.-H. Nguyen, K. X. Nguyen, D. a. Muller, D. C. Ralph, R. a. Buhrman, and C.-F. Pai, Appl. Phys. Lett. **106**, 222402 (2015).

[19] See Supplemental Material

[20] S. V. Aradhya, G. E. Rowlands, J. Oh, D. C. Ralph, and R. A. Buhrman, Nano Lett. **16**, 5987 (2016).

[21] E. B. Myers, F. J. Albert, J. C. Sankey, E. Bonet, R. a Buhrman, and D. C. Ralph, Phys. Rev. Lett. **89**, 196801 (2002).

[22] J. Z. Sun, R. P. Robertazzi, J. Nowak, P. L. Trouilloud, G. Hu, D. W. Abraham, M. C. Gaidis, S. L. Brown, E. J. O'Sullivan, W. J. Gallagher, and D. C. Worledge, Phys. Rev. B **84**, 64413 (2011).

[23] G. E. Rowlands, S. V Aradhya, S. Shi, E. H. Yandel, J. Oh, and D. C. Ralph, Appl. Phys. Lett. **110**, 122402 (2017).

[24] J. Katine, F. Albert, R. Buhrman, E. Myers, and D. Ralph, Phys. Rev. Lett. **84**, 3149 (2000).

[25] J. Sun, Phys. Rev. B **62**, 570 (2000).

[26] O. Ozatay, P. G. Gowtham, K. W. Tan, J. C. Read, K. A. Mkhoyan, M. G. Thomas, G. D. Fuchs, P. M. Braganca, E. M. Ryan, K. V. Thadani, J. Silcox, D. C. Ralph, and R. A. Buhrman, Nat. Mater. **7**, 567 (2008).

[27] C. F. Pai, M. H. Nguyen, C. Belvin, L. H. Vilela-Leão, D. C. Ralph, and R. A. Buhrman, Appl. Phys. Lett. **104**, 82407 (2014).

[28] C.-F. Pai, Y. Ou, L. H. Vilela-Leão, D. C. Ralph, and R. A. Buhrman, Phys. Rev. B **92**, 64426 (2015).

[29] Y. Ou, C.-F. Pai, S. Shi, D. C. Ralph, and R. A. Buhrman, Phys. Rev. B **94**, 140414(R) (2016).

[30] S. Fukami, M. Yamanouchi, H. Honjo, K. Kinoshita, K. Tokutome, S. Miura, S. Ikeda, N. Kasai, and H. Ohno, Appl. Phys. Lett. **102**, 222410 (2013).

FIG. 1

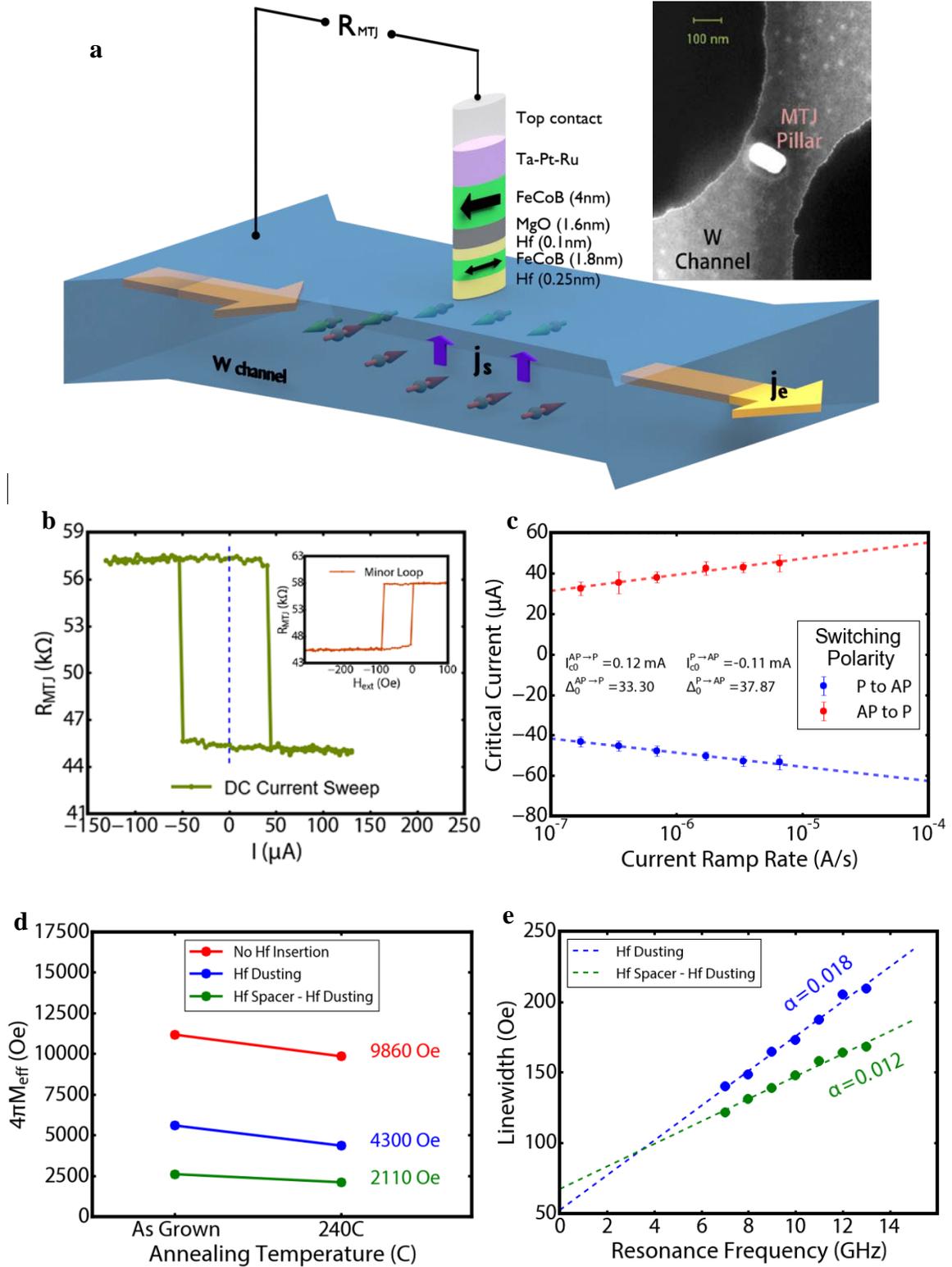

FIG. 2

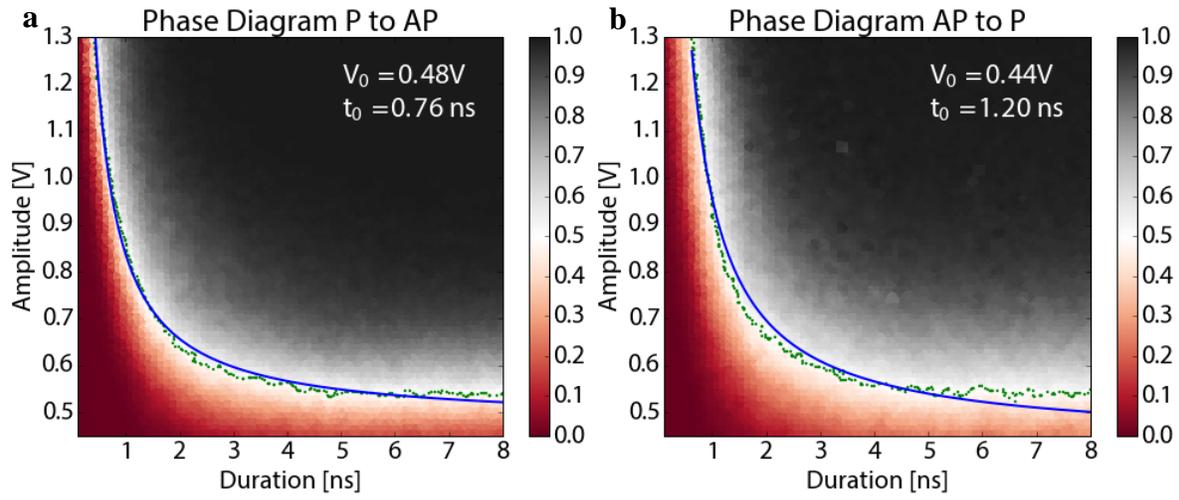

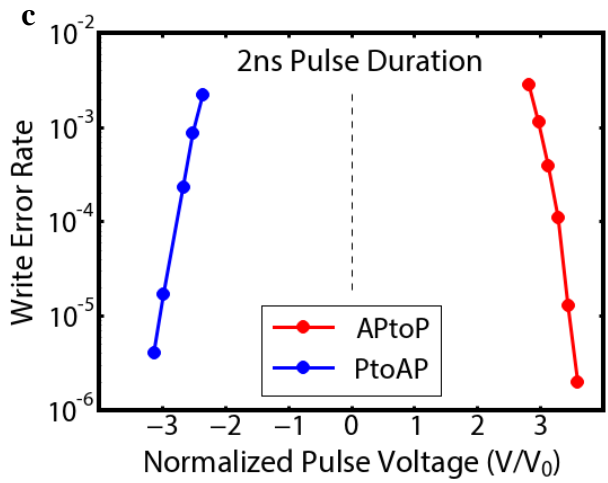

FIG. 3

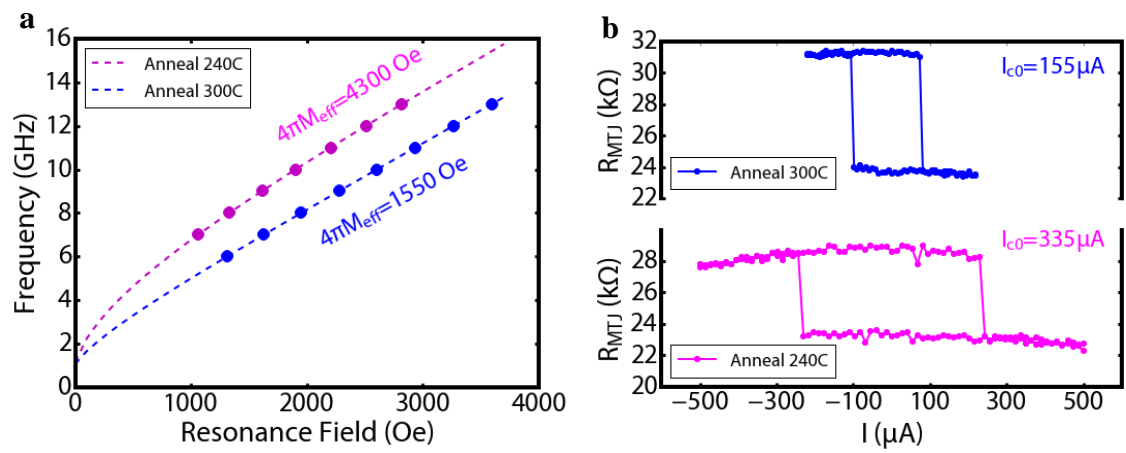

TABLE 1

Comparison of $J_{c0}$ between the different 3T-MTJ structures reported to date.

| | W with Hf insertion layers | W | Pt | Ta | PMA W/CoFeB/MgO nanodot | PMA MTJ with Ta channel | 3T DW motion device |
|---|---|---|---|---|---|---|---|
| Anisotropy | In-plane-magnetized | | | | PMA | | |
| Critical current density (x$10^6$ A/cm$^2$) | 5.4 | 18 | 40 | 32 | >140 | >50 | >50 |
| Reference | This work | Pai et al. [4] | Aradhya et al. [20] | Liu et al. [2] | Fukami et al. [16] | Cubukcu et al. [15] | Fukami et al. [30] |

**FIG. 1** Hf spacer - Hf dusting sample annealed at 240C. (a) The Hf spacer-Hf dusting sample structure and measurement schematics. Inset: SEM image shows a representative nanopillar situated on a W channel after e-beam exposure, development and ion beam etching. (b) Current induced switching loop of the MTJ free layer (FL) showing a thermally assisted switching current of $50\mu A$. The device is 190nm x 30nm and is situated on a 480nm wide W channel. Inset: In-plane field-switching minor loop of the FL. (c) Current ramp rate measurement on the device of (b). Fitting to the macrospin model gives a zero-thermal fluctuation critical current of $115\mu A$ with a thermal stability factor of 35.6. (d) The FL demagnetization field change with annealing temperature for a Hf spacer-Hf dusting sample compared to that of a Hf dusting-only sample and a sample without Hf insertion as measured by flip-chip FMR. $M_{eff}$ significantly decreases in samples with Hf dusting due to enhanced interfacial anisotropy. (e) Linewidths at different resonance frequencies (applied fields) for the Hf dusting-only sample and the Hf spacer-Hf dusting sample measured by flip-chip FMR. Both samples are annealed at 240C. The damping decreases significantly with the insertion of the 0.25nm Hf spacer.

**FIG. 2** Fast and reliable pulse switching of a Hf spacer-Hf dusting sample. Pulse switching phase diagrams and macrospin fits for polarities $P \to AP$ (a) and $AP \to P$ (b) respectively with the switching probability scale bar on the right. Each point is a result of $10^3$ switching attempts. A characteristic switching time of ~1ns and a critical voltage of 0.46V are obtained after fitting 50% probability points (green points) to the macrospin model. (c) WER measurement results for 2 ns square pulses applied to the device (a) and (b). Each point is a result of $10^6$ switching attempts. An error rate as low as approximately $10^{-6}$ is obtained at sufficiently high voltage (current) amplitudes for both polarities.

**FIG. 3** Temperature dependence of the Hf dusting effect. (a) Flip-chip FMR measurement on two Hf dusting-only samples, annealed at 240 C (magenta) and 300 C (blue) respectively, showing a further reduction of $M_{eff}$ at higher annealing temperature.

(b) Current-induced switching of Hf dusting-only samples annealed at two different temperatures, 240 C (magenta) and 300 C (blue). The spin torque switching loops indicate a substantial reduction in critical current with the higher temperature anneal as quantified by the results of ramp rate measurements of $I_{c0}$.